\documentclass[10pt,twocolumn,letterpaper]{article}

\usepackage{iccv}
\makeatletter
\@namedef{ver@everyshi.sty}{}
\makeatother
\usepackage{times}
\usepackage{epsfig}
\usepackage{graphicx}
\usepackage{amsmath}
\usepackage{amssymb}

\usepackage{bm}                  
\usepackage{subcaption}          
\usepackage[export]{adjustbox}   
\usepackage{wrapfig}             
\usepackage{float}               
\usepackage{hyperref}

\usepackage{tikz,pgfplots}       
\usetikzlibrary{spy,calc}        

\newcommand{\img}{\mathbf{x}}
\newcommand{\sysout}{\hat{\img}}
\newcommand{\latent}{\mathbf{y}}
\newcommand{\qlatent}{\hat{\latent}}
\newcommand{\synthparam}{\bm{\theta}}
\newcommand{\armparam}{\bm{\psi}}
\newcommand{\synthmlp}{f_{\synthparam}}
\newcommand{\armmlp}{p_{\armparam}}
\newcommand{\qsteparm}{\Delta_{\armparam}}
\newcommand{\qstepsynth}{\Delta_{\synthparam}}

\definecolor{darkpastelpurple}{rgb}{0.59, 0.44, 0.84}
\definecolor{customred}{HTML}{E76F51}
\definecolor{customblue}{HTML}{376996}
\definecolor{persiangreen}{HTML}{2A9D8F}

\iccvfinalcopy 

\ificcvfinal\fi

\begin{document}

\title{COOL-CHIC: Coordinate-based Low Complexity Hierarchical Image Codec}

\author{Th\'eo Ladune, Pierrick Philippe, F\'elix Henry, Gordon Clare, Thomas Leguay\\
Orange Innovation, France\\
{\tt\small firstname.lastname@orange.com}
}

\maketitle
\ificcvfinal\thispagestyle{empty}\fi

\begin{abstract}
   We introduce COOL-CHIC, a Coordinate-based Low Complexity Hierarchical Image
   Codec. It is a learned alternative to autoencoders with 629 parameters
   and 680 multiplications per decoded pixel. COOL-CHIC offers compression
   performance close to modern conventional MPEG codecs such as HEVC and is
   competitive with popular autoencoder-based systems. This method is inspired
   by Coordinate-based Neural Representations, where an image is represented
   as a learned function which maps pixel coordinates to RGB values. The
   parameters of the mapping function are then sent using entropy coding. At the
   receiver side, the compressed image is obtained by evaluating the mapping
   function for all pixel coordinates. COOL-CHIC implementation is made
   open-source\footnote{\small{\href{https://github.com/Orange-OpenSource/Cool-Chic/}{\texttt{github.com/Orange-OpenSource/Cool-Chic/}}}}.
\end{abstract}

\section{Introduction and related work}

\begin{figure*}[h]
   \centering
   \includegraphics[width=0.8\textwidth]{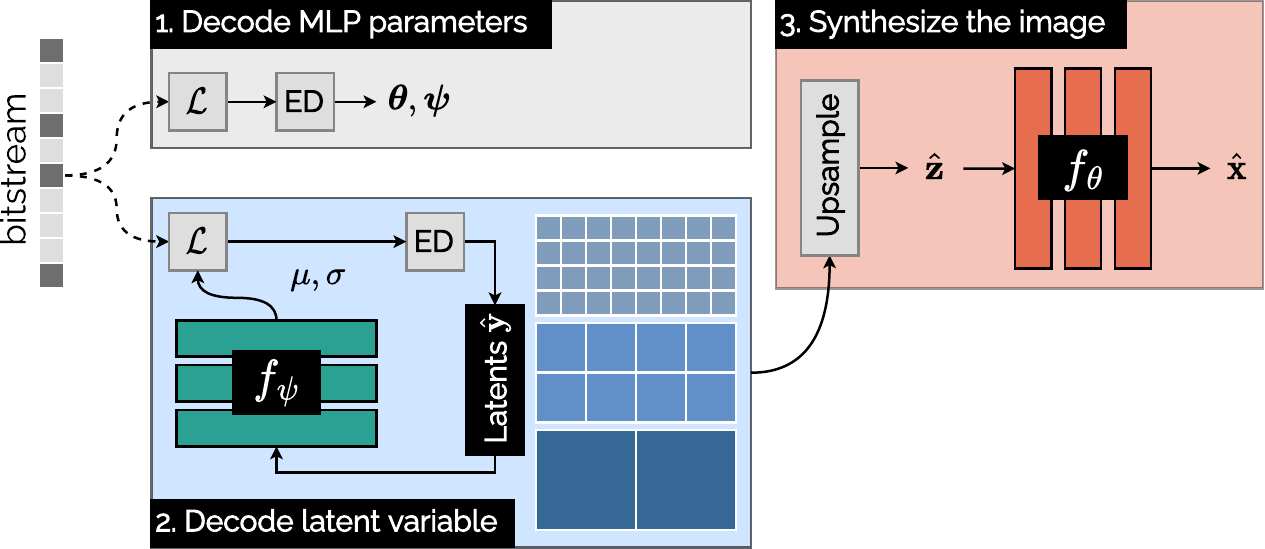}
   \caption{Decoding an image with COOL-CHIC. ED stands for entropy decoding and $\mathcal{L}$ is a Laplace distribution.}
   \label{fig:overview}
\end{figure*}

For years, ITU/MPEG image and video compression algorithms (HEVC \cite{hevc},
VVC \cite{vvc}) have been refining a coding scheme based on the separate
optimization of hand-crafted linear operations. These conventional methods are
being challenged by learned ones, relying on autoencoders to perform a
non-linear mapping from the signal to a compact representation. Ballé's
autoencoder with hyperprior \cite{DBLP:conf/iclr/BalleMSHJ18} is a popular
example of such architecture. These systems are optimized end-to-end on a large
variety of samples. Once the training stage is completed, autoencoders can
generalize to unseen data i.e. they compress all sort of images.

The recent JPEG-AI call for proposal \cite{jpeg-ai-cfp} highlights that learned
image codecs outperform state-of-the-art conventional codecs (VVC). Yet, the
performance of these autoencoders comes at the expense of a tremendous
complexity increase. Indeed, these codecs
\cite{Ma2022a,DBLP:conf/cvpr/HeYPMQW22} have millions of parameters and require
up to a million multiplications to decode a single pixel.
Thus, learned decoders are several orders of magnitude more complex than
conventional ones, which might hinder their adoption.
\newline

In 2021, Dupont introduced COIN \cite{DBLP:journals/corr/abs-2103-03123}, an
image codec relying on Coordinates Neural Representation (CNR) to obtain a
compact representation of an image. COIN models a signal by an overfitted Multi
Layer Perceptron (MLP), performing the mapping from a pixel coordinates to its
RGB value. The encoding stage consists in overfitting the MLP to reconstruct the
image to code. The MLP parameters are then quantized and sent to the receiver.
Finally, performing a forward pass to evaluate the RGB value at each spatial
location allows the receiver to reconstruct the image. Thus, CNR-based codecs
use a lightweight and overfitted decoder, unlike the complex and universal
decoder of auto-encoder approaches. For instance COIN decoder features a
10~000-parameter MLP, with performance close to JPEG.

One of the major limitations of early CNR-based approaches (COIN, NeRF
\cite{mildenhall2020nerf}) is the \textit{non-local} nature of the MLP. Indeed
all parameters of the MLP contribute to the RGB value of all output pixels,
regardless of their position, making the optimization of the parameters
difficult. Recent work such as Instant-NGP \cite{mueller2022instant} or NVP
\cite{DBLP:journals/corr/abs-2210-06823} circumvents this by complementing the
MLP with a latent representation of the image, describing different spatial
locations with different latent parameters.
\newline

This paper introduces COOL-CHIC, a Coordinate-based Low Complexity Hierarchical
Image Codec. It supplements COIN with a hierarchical latent representation,
which contains most of the information about the image. To properly handle the
latent representation, we rely on an auto-regressive model to compress it
through entropy coding. In summary, the contributions of this work are as
follows:
\begin{enumerate}
   \item COOL-CHIC is a learned codec with 680 multiplications per decoded
   pixel. This is orders of magnitude less complex than autoencoders, paving the
   way for learned image decoding without dedicated hardware;
   \item COOL-CHIC offers coding performance close to HEVC and competitive with
   Ballé's hyperprior-based autoencoder, significantly improving COIN results;
   \item COOL-CHIC software is made open-source.
\end{enumerate}

\section{Proposed method}

\subsection{Problem statement and system overview}

Let $\img \in \mathbb{N}^{H \times W \times F}$ be an $H \times W$ image with
$F$ color channels. Following a lossy coding setting, the original image is
allowed to be distorted into $\sysout$ to further decrease its rate. The
objective of lossy coding is to minimize both the rate required to transmit
$\sysout$ and the distortion between $\img$ and $\sysout$. This is stated as the
minimization of the loss function:
\begin{equation}
   \operatorname*{min}\ \mathrm{D}\left(\img, \sysout\right) + \lambda \mathrm{R}\left(\sysout\right),
   \label{eq:rd}
\end{equation}
where $\mathrm{D}$ is a distortion metric (e.g. MSE) and $\mathrm{R}$
denotes the rate in bits per pixel. The rate-control parameter $\lambda \in \mathbb{R}$ balances
the rate-distortion tradeoff.
\newline

Decoding an image with COOL-CHIC consists of three main steps, shown in Fig.
\ref{fig:overview}. First, the parameters of two MLPs ($f_{\armparam}$ and
$f_{\synthparam}$) are retrieved from the bitstream. Then, $f_{\armparam}$ is
used to decode $\qlatent$, a set of $L$ 2-dimensional discrete latent variables.
The latent variables are upsampled and concatenated as a dense 3D representation
$\hat{\mathbf{z}} = \mathtt{upsample}(\qlatent)$. Finally, the RGB value of each
pixel $\sysout_{ij}$ from the compressed image is computed by feeding the dense
latent representation to the synthesis MLP:
\begin{equation}
   \sysout_{ij} = \synthmlp(\hat{\mathbf{z}}_{ij}).
\end{equation}

Encoding an image is achieved by overfitting the parameters $\left\{\qlatent,
\armparam, \synthparam \right\}$ so that they minimize the rate-distortion cost
of the image stated in eq. \eqref{eq:rd}. These parameters are learned for a
single image and are therefore adapted for this content. Thus, the parameters
$\left\{\qlatent, \armparam, \synthparam \right\}$ are the compressed
representation of the image $\img$. Efficient transmission of these parameters
uses entropy coding, which relies on an estimate $p$ of the signal's (unknown)
probability distribution $q$. Asymptotically, entropy coding algorithms offer a
rate close to the cross-entropy of the signal:
\begin{equation}
   R(\qlatent) = \mathbb{E}_{\qlatent \sim q} \left[-\log_2 p(\qlatent)\right].
   \label{eq:rate1}
\end{equation}

Here, special attention is dedicated to the latent distribution $p(\qlatent)$,
as its dimension is several orders of magnitude larger than that of the MLP
parameters $\synthparam$ and $\armparam$. To this end, a lightweight
auto-regressive probability model $\armmlp \left(\qlatent\right)$ is implemented
by an MLP $f_{\armparam}$. COOL-CHIC strives to minimize the RD cost expressed
in eq. \eqref{eq:rd}, rewritten to more clearly expose the optimized quantities:
\begin{equation}
   \operatorname*{min}_{\qlatent, \synthparam, \armparam} \
   \mathrm{D}\left(\img, \synthmlp(\mathtt{upsample}(\qlatent))\right) - \lambda \log_2 \armmlp \left(\qlatent\right).
   \label{eq:rd2}
\end{equation}
This objective function does not account for the rate associated to the MLP
parameters $\left\{\synthparam, \armparam \right\}$, since they only contribute
marginally to the overall rate. Entropy coding of the MLP parameters is achieved
with a \textit{non}-learned distribution, estimated once the optimization is
complete.
\newline

In the following, Section \ref{subsec:synthesis} presents the synthesis module
which reconstructs the decoded image $\sysout$ from the latent variables.
Section \ref{subsec:arm} introduces the auto-regressive module which models the
distribution $\armmlp(\qlatent)$. Finally, Section \ref{subsec:mlp_coding}
describes how the MLP parameters are transmitted.

\subsection{Synthesis module}
\label{subsec:synthesis}

\begin{figure*}
   \centering
   \includegraphics[width=0.9\textwidth]{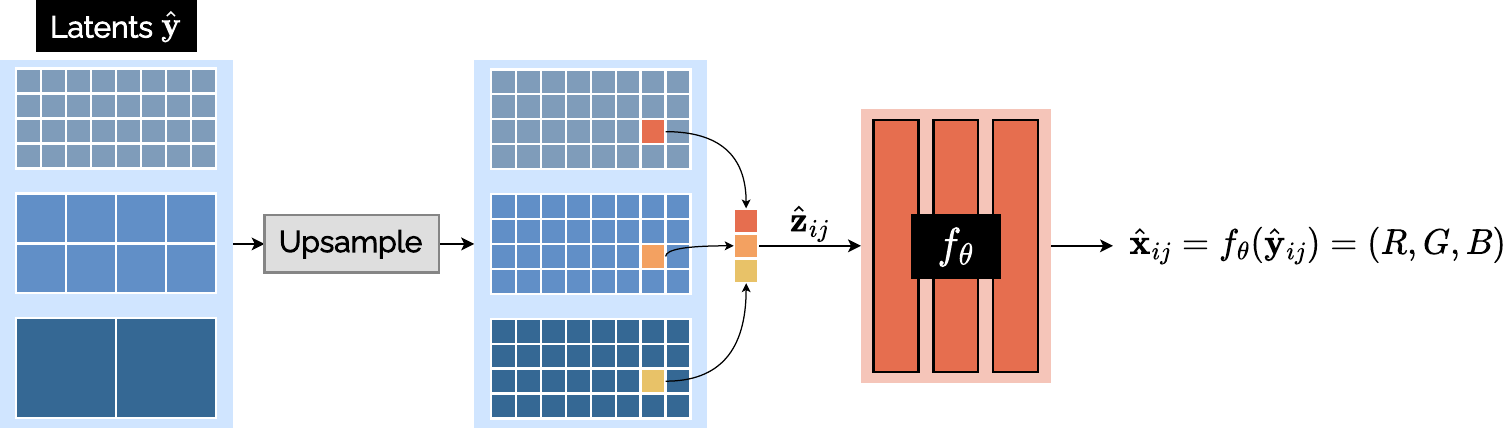}
   \caption{COOL-CHIC synthesis illustrated for a $4 \times 8$ image with
   $L = 3$ latent channels.}
   \label{fig:synthesis}
\end{figure*}

COOL-CHIC synthesis is depicted in Fig. \ref{fig:synthesis}.
Inspired by recent CNR approaches, it relies on a latent representation
$\qlatent$ fed to a lightweight MLP $\synthmlp$. This latent representation
consists of $L$ channels of different spatial resolutions:
\begin{align}
   \qlatent &= \left\{ \qlatent_{k} \in \mathbb{Z}^{H_k \times W_k},\ k = 0, \ldots, L - 1\right\},\\
   &\textrm{ with } H_k = \frac{H}{2^k} \textrm{ and } W_k = \frac{W}{2^k}. \nonumber
\end{align}

The hierarchical nature of $\qlatent$ allows for a compact representation of
low-frequency areas thanks to the lowest resolutions of $\qlatent$, while still
being able to capture fine details on the highest resolutions. Typically, $L =
7$ is used so that the coarsest latent resolution is $\frac{H}{64} \times \frac{W}{64}$.
\newline

Each channel $\qlatent_{k}$ of the latent representation is then upsampled by a
factor $2^k$ with a bicubic interpolation. The upsampled latents are concatenated
along the channel dimension to obtain
$\hat{\mathbf{z}}~=~\mathtt{upsample}(\qlatent)$, a dense 3D representation of size
$H \times W \times L$. Finally, the RGB value of each pixel $\sysout_{ij}$ from
the compressed image is computed by sampling $\hat{\mathbf{z}}$ at the desired
position and feeding the resulting $L$-dimensional vector to the synthesis MLP:
\begin{equation}
   \sysout_{ij} = \synthmlp(\hat{\mathbf{z}}_{ij}), \text{ with }\hat{\mathbf{z}}_{ij} = \left\{\hat{z}_{ijk}, k = 0, \ldots, L - 1\right\}.
\end{equation}


The whole synthesis module ($\synthmlp$ and $\qlatent$), is obtained by gradient
descent to minimize the rate-distortion cost stated in eq \eqref{eq:rd2}.
However, the latent representation is required to be discrete in order to be
entropy coded. As it is not possible to directly optimize a discrete variable
through a gradient descent-based optimization, a continuous latent
representation $\latent$ is learnt as a proxy. Moreover, the quantization
operation is replaced by noise addition during training
\cite{DBLP:conf/iclr/BalleLS17}:

\begin{equation}
   \qlatent = \left\{
   \begin{array}{lll}
      \latent + \mathbf{u}&\textrm{ with } \mathbf{u} \sim \mathcal{U}\left[-0.5, 0.5\right] & \textrm{if learning } \latent,\\
      Q(\latent)&\textrm{ with } Q \textrm{ a uniform quantizer} & \textrm{otherwise.}%
   \end{array}
   \right.
\end{equation}

\subsection{Auto-regressive probability model}
\label{subsec:arm}

\begin{figure*}
   \centering
   \includegraphics[width=0.7\textwidth]{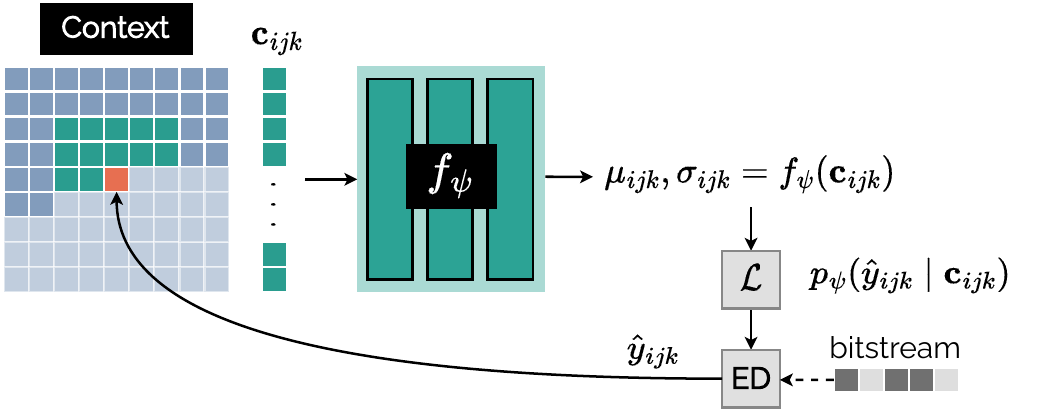}
   \caption{Entropy decoding of $\hat{y}_{ijk}$ using the auto-regressive probability model of COOL-CHIC and $C = 12$ context pixels.}
   \label{fig:arm}
\end{figure*}

The role of the auto-regressive probability model $\armmlp$ is highlighted by
rewriting eq. \eqref{eq:rate1}:
\begin{align}
   R(\qlatent)
   &= \mathbb{E}_{\qlatent \sim q} \left[-\log_2 p(\qlatent)\right] \nonumber \\
   &= \mathbb{E}_{\qlatent \sim q} \left[-\log_2 \armmlp(\qlatent) + \log_2 q(\qlatent) - \log_2 q(\qlatent) \right] \nonumber\\
   &= D_{KL} \left(q\ ||\ \armmlp\right) + \mathrm{H}(\qlatent). \label{eq:rate2}
\end{align}

Here, $D_{KL}$ stands for the Kullback-Leibler divergence and $\mathrm{H}$ for
Shannon's entropy. Equation \eqref{eq:rate2} states that it is possible to act
on two terms to minimize the rate. First, decreasing the entropy (i.e.
the average information quantity) of the latent variable $\qlatent$. This is at
the heart of the rate-distortion tradeoff as less information in $\qlatent$
implies more distortion in $\sysout$. The second means of reducing the rate is to
estimate a distribution $\armmlp$ as close as possible to the actual (unknown)
latent distribution $q$. This is the role of the auto-regressive module shown in
Fig. \ref{fig:arm}.
\newline

Modeling the joint distribution of $\qlatent$ is untractable due to its high
dimension. Inspired by \cite{DBLP:journals/corr/abs-1809-02736},
we resort to a factorized model, where the distribution of each latent pixel
$\hat{y}_{ijk}$ (i.e. the pixel at location $(i, j)$ in the $k$-th latent
channel) is conditioned on $C$ spatially neighboring pixels $\mathbf{c}_{ijk}
\in \mathbb{Z}^C$.
\begin{equation}
   \armmlp(\qlatent) = \prod_{i,j,k} \armmlp(\hat{y}_{ijk} \mid \mathbf{c}_{ijk}).
\end{equation}

Since the distribution $\armmlp$ must be known to both the emitter and receiver,
only causal (already received) context pixels can be used to estimate the
distribution $\armmlp$. Moreover, the context pixels are selected to introduce
as little sequentiality as possible. As such, no inter latent channel dependency
is leveraged, allowing to decode all $L$ channels in parallel. Furthermore, rows
can also be processed in parallel in a wavefront-like manner \cite{6327343}.
\newline

Following the usual practice in learned coding
\cite{DBLP:conf/iclr/BalleMSHJ18}, the discrete distribution $\armmlp(\qlatent)$
of the quantized latent variable is actually modeled by integrating the
\textit{continuous} distribution of the non-quantized latent $g(\latent)$,
modeled as a Laplace distribution. The MLP $f_{\armparam}$ learns to estimate
the proper expectation and scale parameters of $g$, based on the context pixels.
As such, the probability of a latent pixel is:
\begin{align}
   &\armmlp(\hat{y}_{ijk} \mid \mathbf{c}_{ijk}) = \int_{\hat{y}_{ijk} - 0.5}^{\hat{y}_{ijk} + 0.5} g(y)\mathrm{d}y
   ,\\
   &\text{ with } g \sim \mathcal{L}\left(\mu_{ijk}, \sigma_{ijk}\right)
   \text{ and }\ \mu_{ijk}, \sigma_{ijk} = f_{\armparam}(\mathbf{c}_{ijk}). \nonumber
   \label{eq:integrate_cdf}
\end{align}
Finally, the rate term present in eq. \eqref{eq:rd2} sums up to:
\begin{align}
   R(\qlatent)
   &= -\log_2 \armmlp(\qlatent) \nonumber\\
   & = -\log_2 \prod_{i,j,k} \armmlp(\hat{y}_{ijk} \mid \mathbf{c}_{ijk}) \nonumber\\
   & = \sum_{i,j,k} -\log_2 \armmlp(\hat{y}_{ijk} \mid \mathbf{c}_{ijk}).
\end{align}

\subsection{Compressing the model parameters}
\label{subsec:mlp_coding}

During the training stage (i.e. the encoding), the MLPs parameters
$\left\{\armparam, \synthparam\right\}$ are represented as 32-bit floating point
values. Yet, they do not require such a high-precision
representation once the training is finished. This section explains how
the quantization accuracy is set for the MLPs.
\newline



As the synthesis $\synthmlp$ and the probability model $\armmlp$ perform
different tasks, they likely require different accuracy. Consequently, different
quantization steps $\qsteparm$ and $\qstepsynth$ are used for $\armparam$ and
$\synthparam$. Instead of the full-precision parameters determined by the
optimization process, COOL-CHIC relies on their quantized version:
\begin{align}
   \hat{\synthparam} &= Q(\synthparam, \qstepsynth) \text{ and }\
   \hat{\armparam} = Q(\armparam, \qsteparm), \nonumber \\
   &\text{ with } Q(\cdot, \Delta) \text{ a scalar quantizer of step } \Delta.
\end{align}

A probability model of $\hat{\synthparam}$ and $\hat{\armparam}$ is required to
send them \textit{via} an entropy coding algorithm. Similarly to the quantized
latent variable distribution, the discrete distribution of each quantized MLP
parameter is modeled \textit{via} a continuous Laplace distribution, see eq.
\eqref{eq:integrate_cdf}. As such, the probability of one parameter from
$\synthparam$ (the same holds for $\armparam$) is:
\begin{align}
   p(\hat{\theta}_i) &= \int_{\hat{\theta}_i - 0.5}^{\hat{\theta}_i + 0.5}g(\theta)\mathrm{d}\theta, \nonumber \\
   &\text{ with } g \sim \mathcal{L}\left(0, \sigma_{\hat{\synthparam}}\right)
   \text{ and } \sigma_{\hat{\synthparam}} = \mathtt{stddev}(\hat{\synthparam})
\end{align}
The total rate contribution of both MLPs is estimated as:
\begin{align}
   \mathrm{R}_{\text{MLP}} &= R_{\hat{\synthparam}} + R_{\hat{\armparam}} \nonumber \\
      &= \sum_{\hat{\theta}_i} -\log_2 p(\hat{\theta}_i) + \sum_{\hat{\psi}_j} -\log_2 p(\hat{\psi}_j).
\end{align}

Quantization of the MLP parameters allows the reduction of
$\mathrm{R}_{\text{MLP}}$ at the expense of the probability model and synthesis
accuracy. Consequently, it is important to properly select the value of
$\qsteparm$ and $\qstepsynth$. This is achieved by a greedy minimization of the
rate-distortion cost associated to different quantization steps (e.g. from
$10^{-1}$ to $10^{-5}$). For each quantization step, $\mathrm{R}_{\text{MLP}}$
and the compression performance (distortion and latent rate when using the
quantized MLPs) are measured. The selected quantization steps are those
minimizing the following rate-distortion cost.
\begin{equation}
   \operatorname*{min}_{\qsteparm, \qstepsynth} \
   \mathrm{D}\left(\img, \hat{\img}\right) + \lambda \left(
      \mathrm{R}(\qlatent) + \mathrm{R}_{\text{MLP}}
   \right).
\end{equation}

\section{Experimental results}

\subsection{Rate-distortion results}
\label{sec:rd_results}

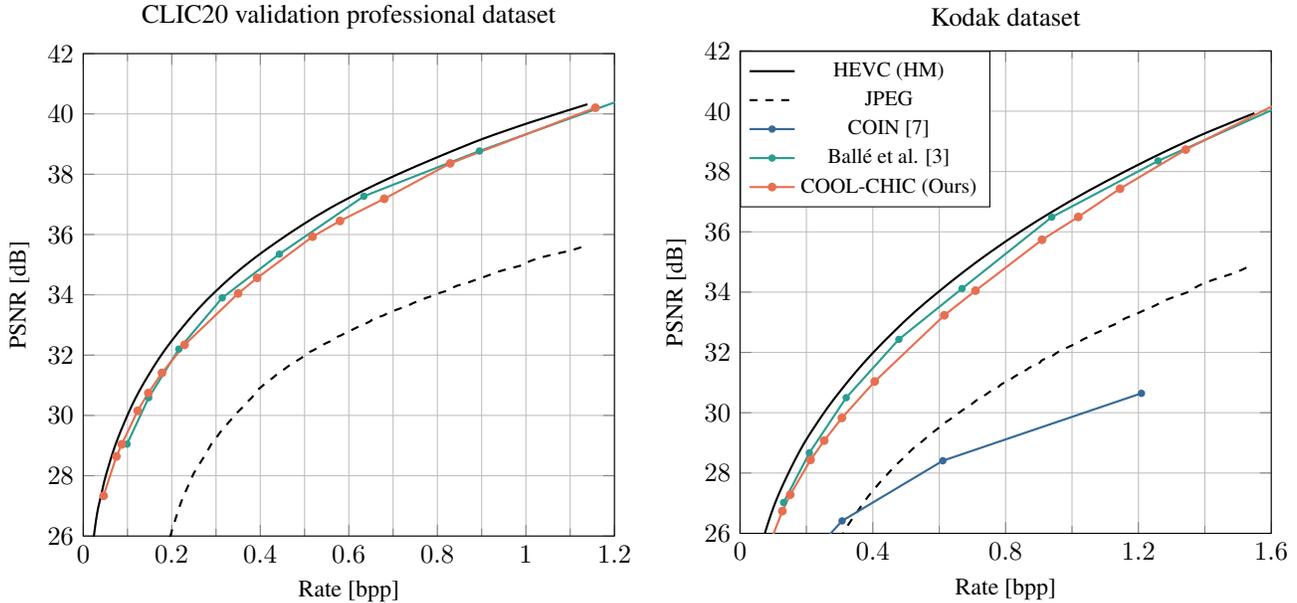
\begin{figure*}
   \centering
   \begin{subfigure}{0.495\textwidth}
      \centering
      \begin{tikzpicture}
         \begin{axis}[
            grid= both,
            xlabel = {\small Rate [bpp]},
            ylabel = {\small PSNR [dB]} ,
            xmin = 0, xmax = 1.2,
            ymin = 26, ymax = 42,
            ylabel near ticks,
            xlabel near ticks,
            height=8cm,
            width=\linewidth,
            xtick distance={0.2},
            ytick distance={2},
            minor y tick num=0,
            minor x tick num=1,
            title={CLIC20 validation professional dataset}
         ]


            \addplot[solid, thick, black, smooth] table [x=rate_bpp,y=psnr] {data/clic20pro-valid/hm.txt};

            \addplot[dashed, thick, black, smooth] table [x=rate_bpp,y=psnr] {data/clic20pro-valid/jpeg.txt};

            \addplot[solid, thick, persiangreen, mark=*,mark size=1pt] table [x=rate_bpp,y=psnr] {data/clic20pro-valid/balle18_hyperprior_mse.txt};

            \addplot[solid, thick, customred, mark=*, mark size=1.25pt] table [x=rate_bpp,y=psnr] {data/clic20pro-valid/12t_12_mse.txt};



         \end{axis}
      \end{tikzpicture}
   \end{subfigure}
   \begin{subfigure}{0.495\textwidth}
   \centering
   \begin{tikzpicture}
      \begin{axis}[
         grid= both,
         xlabel = {\small Rate [bpp]},
         ylabel = {\small PSNR [dB]} ,
         xmin = 0, xmax = 1.6,
         ymin = 26, ymax = 42,
         ylabel near ticks,
         xlabel near ticks,
         height=8cm,
         width=\linewidth,
         xtick distance={0.4},
         ytick distance={2},
         minor y tick num=0,
         minor x tick num=1,
         legend style={at={(0, 1)}, anchor=north west},
         title={Kodak dataset}
      ]


         \addplot[solid, thick, black, smooth] table [x=rate_bpp,y=psnr] {data/kodak/hm.txt};
         \addlegendentry{\footnotesize HEVC (HM)}

         \addplot[dashed, thick, black, smooth] table [x=rate_bpp,y=psnr] {data/kodak/jpeg.txt};
         \addlegendentry{\footnotesize JPEG}

         \addplot[solid, thick, customblue, mark=*, mark size=1pt] table [x=rate,y=psnr] {data/kodak/coin.txt};
         \addlegendentry{\footnotesize COIN \cite{DBLP:journals/corr/abs-2103-03123}}



         \addplot[thick, thick, persiangreen, mark options={solid}, mark=*, mark size=1pt] table [x=rate_bpp,y=psnr] {data/kodak/balle18_hyperprior_mse.txt};
         \addlegendentry{\footnotesize Ballé et al. \cite{DBLP:conf/iclr/BalleMSHJ18}}


         \addplot[solid, thick, customred, mark=*, mark size=1.25pt] table [x=rate_bpp,y=psnr] {data/kodak/12t_12_mse.txt};
         \addlegendentry{\footnotesize COOL-CHIC (Ours)}


      \end{axis}
   \end{tikzpicture}
\end{subfigure}
\caption{Rate-distortion performance on CLIC20 professional and Kodak datasets.
Performance of Ballé's autoencoder come from CompressAI
\cite{begaint2020compressai}. PSNR is computed in the RGB444 domain.}
\label{fig:rd_results}
\end{figure*}

This section provides experimental results demonstrating the efficiency of
COOL-CHIC as a low-complexity image decoder. To this end, the rate-distortion
performance of COOL-CHIC is measured on the Kodak dataset \cite{kodak} and the
CLIC20 professional validation set \cite{clic20pro} using the PSNR as quality
metric (computed in the RGB444 domain).

Several anchors are provided to better appreciate COOL-CHIC results. A first set
of anchors are conventional (i.e. non-learned) codecs such as JPEG and HEVC
intra, where HEVC is tested using its test model HM operating in 444. The
performance of COIN \cite{DBLP:journals/corr/abs-2201-12904}, a popular
CNR-based codec from the literature, is provided. Finally, results of the
well-known Ballé's autoencoder with hyperprior \cite{DBLP:conf/iclr/BalleMSHJ18}
(inferred using compressAI \cite{begaint2020compressai}) are also presented.
\newline

Prior CNR-based codecs (e.g. COIN) rely on varying the architecture of the
synthesis MLP $\synthmlp$ to address different rates. Thanks to the addition of
a latent representation, COOL-CHIC implements a single architecture for all
rates. Both the synthesis MLP $\synthmlp$ and the probability model MLP
$f_{\armparam}$ have two hidden layers of width $12$ with ReLUs. As stated in
Fig. \ref{fig:arm}, $C = 12$ context pixels are leveraged and the latent
variable is composed of $L = 7$ different channels. This results in a
lightweight decoder, with 629 parameters and 680 multiplications per decoded
pixel.
\newline


Figure \ref{fig:rd_results} presents the rate-distortion results. COOL-CHIC
outperforms prior CNR-based codec COIN, across the entire range of rate.
Moreover, it offers performance on par with Ballé's autoencoder while offering a
reduced decoding complexity. At higher rates, COOL-CHIC comes close to the
performance of modern conventional codecs such as HEVC. These are compelling
results since they prove that COOL-CHIC is able to compete with well-established
autoencoders and conventional codecs.

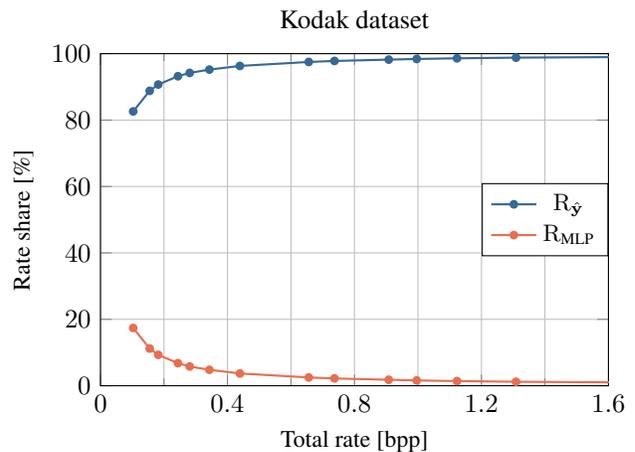
\begin{figure}[b]
   \centering
   \begin{tikzpicture}
      \begin{axis}[
         grid= both,
         xlabel = {\small Total rate [bpp]},
         ylabel = {\small Rate share [\%]} ,
         xmin = 0, xmax = 1.6,
         ymin = 0, ymax = 100,
         ylabel near ticks,
         xlabel near ticks,
         height=6cm,
         width=\linewidth,
         xtick distance={0.4},
         ytick distance={20},
         minor y tick num=0,
         minor x tick num=1,
         legend style={at={(1,0.5)}, anchor=east},
         title={Kodak dataset},
      ]

      \addplot[solid, thick, customblue, mark=*, mark size=1.25pt] table [x=rate,y expr={100 - \thisrow{share_rate_mlp}}] {data/kodak/rate_mlp.txt};
      \addlegendentry{\small $\mathrm{R}_{\qlatent}$}

      \addplot[solid, thick, customred, mark=*, mark size=1.25pt] table [x=rate,y=share_rate_mlp] {data/kodak/rate_mlp.txt};
      \addlegendentry{\small $\mathrm{R}_{\text{MLP}}$}
      \end{axis}
   \end{tikzpicture}
   \caption{Contributions of the two rate terms: the MLPs rate
   $\mathrm{R}_{\text{MLP}}$ and the latent variable rate $\mathrm{R}_{\qlatent}$.}
   \label{fig:rate_mlp}
\end{figure}

It should be noted that COOL-CHIC performance appears to be better for higher
rates. This is likely due to the cost of sending the MLPs for the synthesis and
probability model. Figure \ref{fig:rate_mlp} presents the evolution of the rate
share for the MLPs and the latent variable. At lower rates, the rate associated
to the MLPs approaches 20~\% of the overall rate. This is an important overhead
which might explains the worse performance of COOL-CHIC at low rates. As the
MLPs rate is approximately constant for all rates, this overhead tends to
decrease when the overall bitrate increases.

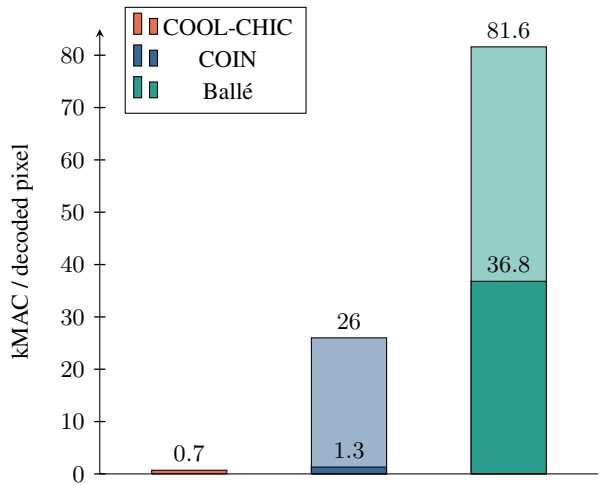
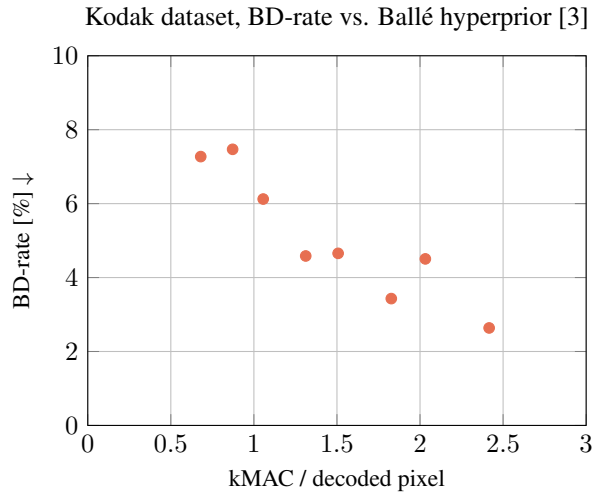
\begin{figure*}
   \centering
   \begin{subfigure}{0.47\textwidth}
      \centering
      \begin{tikzpicture}
         \begin{axis}[
            height=7.5cm,
            width=\linewidth,
            ylabel={kMAC / decoded pixel},
            xmin=0.5, xmax=3.5,
            ymin=0, ymax=85,
            ytick distance=10,
            minor y tick num=0,
            enlarge x limits=0.02,
            xticklabels={},
            xtick={0,...,3},
            xtick pos=left,
            ytick pos=left,
            axis x line*=center,
            axis y line=left,
            xlabel style={right},
            ylabel style={yshift=-.2cm},
            bar width=1cm,                       
            ybar=-1cm,                           
            xtick style={draw=none},
            ytick style={black, line width=0.5pt},
            axis on top,
            nodes near coords,
            x tick label style={rotate=45,anchor=east, align=center},
            every axis/.append style={font=\small},
            clip=false,
            legend entries={COOL-CHIC,COIN,Ballé},
            legend style={at={(0.05,1.05)},anchor= north west},
            title style={yshift=-0.75ex},
            unbounded coords=jump,
         ]
            \addplot[ybar, fill=customred] coordinates {(1,0.7)};
            \addplot[ybar, fill=customblue!50, forget plot] coordinates {(2,26)};
            \addplot[ybar, fill=customblue] coordinates {(2,1.3)};
            \addplot[ybar, fill=persiangreen!50, forget plot] coordinates {(3,81.6)};
            \addplot[ybar, fill=persiangreen] coordinates {(3,36.8)};
      \end{axis}
   \end{tikzpicture}
   \caption{Learned decoders minimum and maximum complexity.}
   \label{fig:complexity_learned_decoders}
   \end{subfigure}
   \begin{subfigure}{0.47\textwidth}
      \centering
      \begin{tikzpicture}
         \begin{axis}[
            grid= both,
            xlabel = {\small kMAC / decoded pixel},
            ylabel = {\small BD-rate [\%] $\downarrow$} ,
            xmin = 0, xmax = 3,
            ymin = 0, ymax = 10,
            ylabel near ticks,
            xlabel near ticks,
            height=6.5cm,
            width=\linewidth,
            xtick distance={0.5},
            ytick distance={2},
            minor y tick num=0,
            minor x tick num=0,
            legend style={at={(1,0.)}, anchor=south east},
            title={Kodak dataset, BD-rate vs. Ballé hyperprior \cite{DBLP:conf/iclr/BalleMSHJ18}},
         ]
            \addplot[dashed, only marks, customred, mark=*, mark options={solid}] table [x expr={0.001 * \thisrow{mac}},y=bdrate] {data/complexity/kodak_bdrate_mac.txt};
         \end{axis}
      \end{tikzpicture}
      \caption{Performance for different decoding complexity. Lower BD-rate is better}
      \label{fig:bdrate_mac}
      
   \end{subfigure}
   \caption{Complexity analysis of COOL-CHIC decoder. kMAC stands for kilo multiplication-accumulation.}
\end{figure*}

\subsection{Complexity analysis}
\label{sec:results_complexity}

\subsubsection{Decoder complexity}

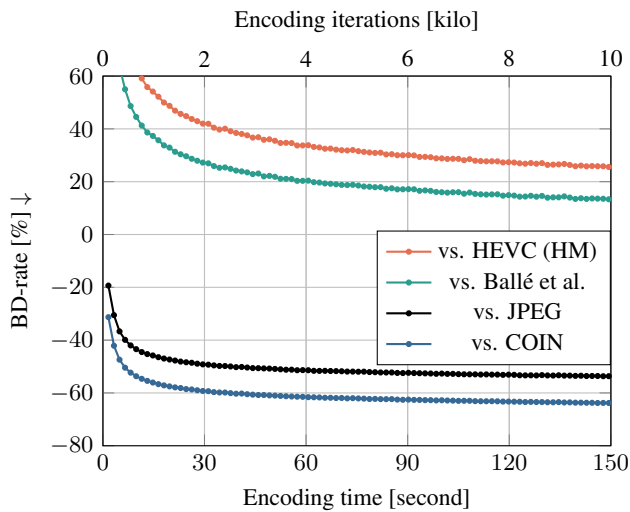
\begin{figure}[b]
   \centering
   \begin{tikzpicture}
      \pgfplotsset{
         compat=1.16,
         EncodingTime plot style/.style={
            every tick label/.append style={font=\small},
            xmin = 0, xmax = 150,
            ymin = -80, ymax = 60,
            ylabel near ticks,
            xlabel near ticks,
            height=6.5cm,
            width=\linewidth,
            xtick distance={30},
            ytick distance={20},
            grid= both,
            domain=\pgfkeysvalueof{/pgfplots/xmin}:\pgfkeysvalueof{/pgfplots/xmax},
         },
      }
      \begin{axis}[
         EncodingTime plot style,
         xlabel={\small Encoding iterations [kilo]},
         axis x line*=right,
         xticklabels={, 0, 2, 4, 6, 8, 10, 12, 14, 16},  
         ytick=\empty,
      ]
      \addplot[
         draw=none, 
         samples=2 
         ]{0};
      \end{axis}
      \begin{axis}[
         EncodingTime plot style,
         xlabel = {\small Encoding time [second]},
         ylabel = {\small BD-rate [\%] $\downarrow$} ,
         legend style={at={(1,0.4)}, anchor=east},
      ]

         \addplot[solid, mark=*, thick, customred, mark size=0.75pt] table [x=time_second,y=bd_rate] {data/encoding_time/bd_rate_vs_hm.txt};
         \addlegendentry{\small vs. HEVC (HM)}

         \addplot[solid, mark=*, thick, persiangreen, mark size=0.75pt] table [x=time_second,y=bd_rate] {data/encoding_time/bd_rate_vs_balle.txt};
         \addlegendentry{\small vs. Ballé et al.}

         \addplot[solid, mark=*, thick, black, mark size=0.75pt] table [x=time_second,y=bd_rate] {data/encoding_time/bd_rate_vs_jpeg.txt};
         \addlegendentry{\small vs. JPEG}

         \addplot[solid, mark=*, thick, customblue, mark size=0.75pt] table [x=time_second,y=bd_rate] {data/encoding_time/bd_rate_vs_coin.txt};
         \addlegendentry{\small vs. COIN}

      \end{axis}
   \end{tikzpicture}
   \caption{COOL-CHIC BD-rate on the Kodak dataset as a function of the encoding
   time. Negative BD-rate means that COOL-CHIC requires less rate to offer the
   same quality.}
   \label{fig:encoding_time_bd_rate}
\end{figure}

\newcommand{\subfigurewidth}{0.192\linewidth}  
\begin{figure*}[h]
   \centering
   \begin{subfigure}{\subfigurewidth}
      \begin{tikzpicture}[spy using outlines={customred,magnification=4,size=1.5cm}]
         \node {\includegraphics[width=\textwidth]{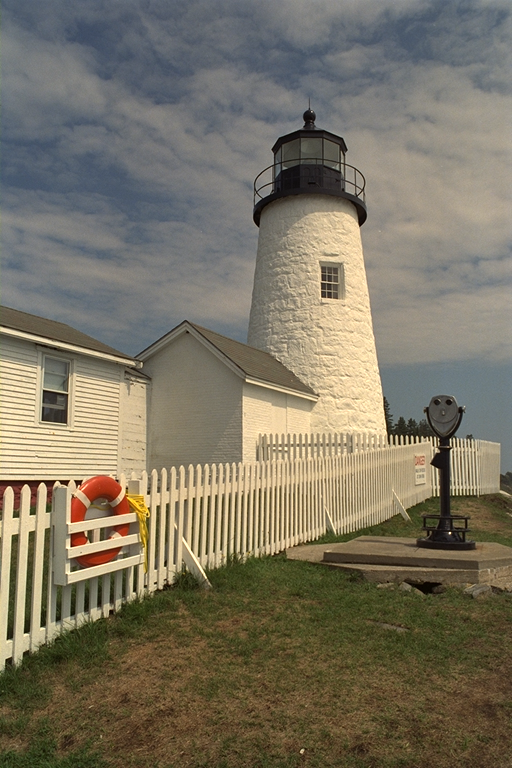}};
         \spy on (0.15,0.1) in node [left] at (0.22,1.177);
      \end{tikzpicture}
      \caption{Original}
   \end{subfigure}
   \begin{subfigure}{\subfigurewidth}
      \begin{tikzpicture}[spy using outlines={customred,magnification=4,size=1.5cm}]
         \node {\includegraphics[width=\textwidth]{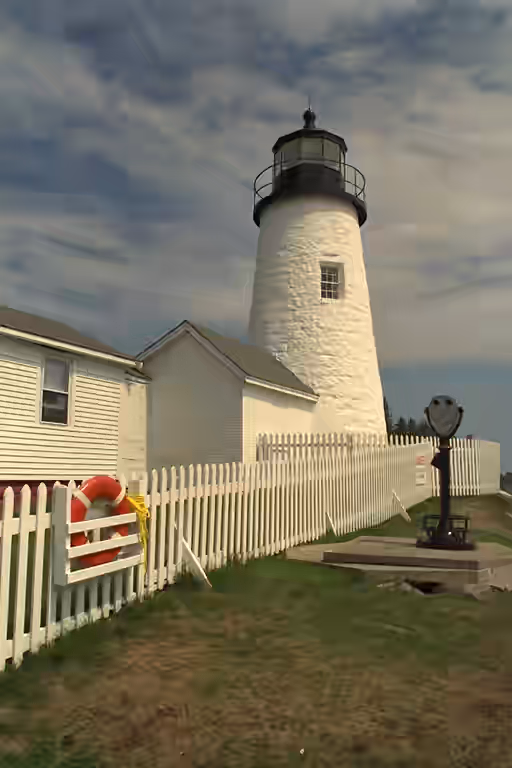}};
         \spy on (0.15,0.1) in node [left] at (0.22,1.177);
      \end{tikzpicture}
      \caption{HEVC 0.142 bpp}  
   \end{subfigure}
   \begin{subfigure}{\subfigurewidth}
      \begin{tikzpicture}[spy using outlines={customred,magnification=4,size=1.5cm}]
         \node {\includegraphics[width=\textwidth]{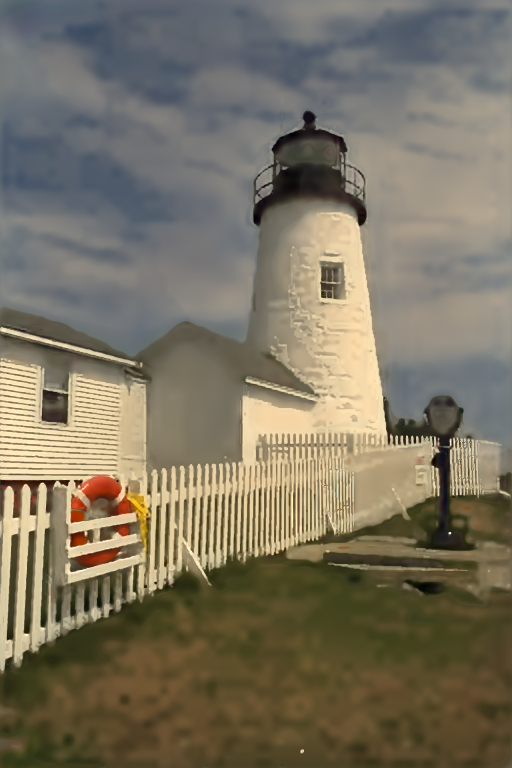}};
         \spy on (0.15,0.1) in node [left] at (0.22,1.177);
      \end{tikzpicture}
      \caption{Ours 0.145 bpp}  
      \label{subfig:visualcomparison_low_rate}
   \end{subfigure}
   \begin{subfigure}{\subfigurewidth}
      \begin{tikzpicture}[spy using outlines={customred,magnification=4,size=1.5cm}]
         \node {\includegraphics[width=\textwidth]{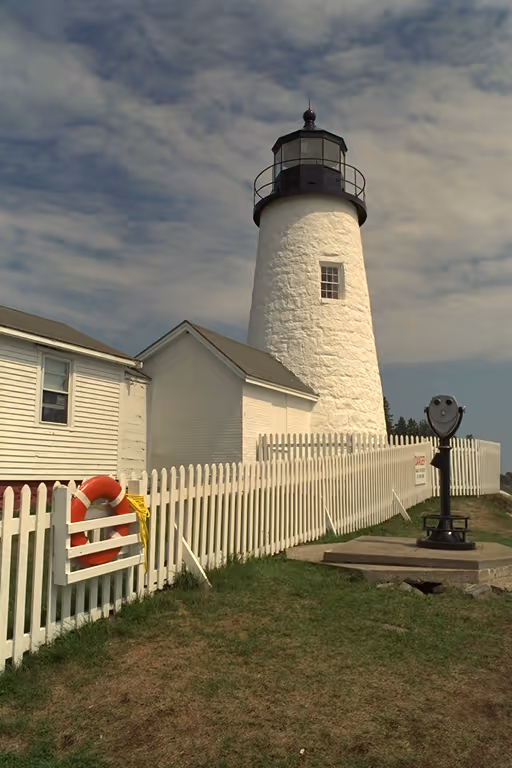}};
         \spy on (0.15,0.1) in node [left] at (0.22,1.177);
      \end{tikzpicture}
      \caption{HEVC 0.578 bpp}   
   \end{subfigure}
   \begin{subfigure}{\subfigurewidth}
      \begin{tikzpicture}[spy using outlines={customred,magnification=4,size=1.5cm}]
         \node {\includegraphics[width=\textwidth]{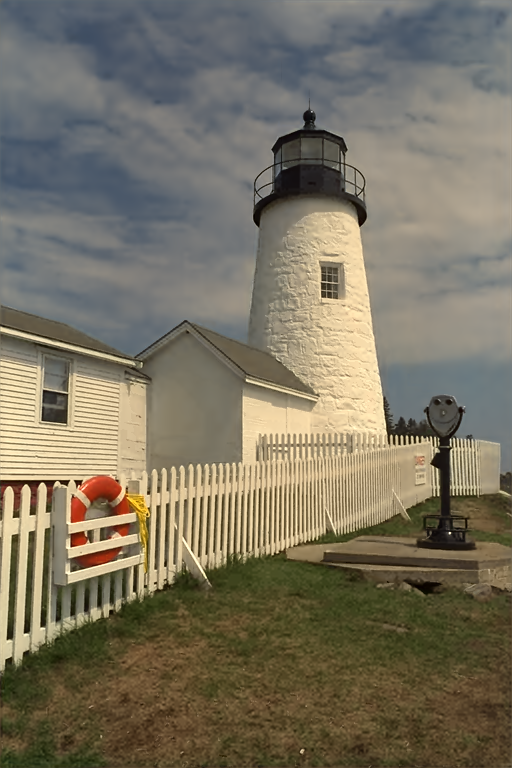}};
         \spy on (0.15,0.1) in node [left] at (0.22,1.177);
      \end{tikzpicture}
      \caption{Ours 0.589 bpp}  
      \label{subfig:visualcomparison_high_rate}
   \end{subfigure}
   \caption{Comparison of HEVC and COOL-CHIC on Kodak image \textit{kodim19} at
   two different rates.}
   \label{fig:visual_comparison}
\end{figure*}

\newcommand{\trimleft}{0.25\width}
\newcommand{\trimright}{0.28\width}
\newcommand{\trimtop}{.04\height}
\newcommand{\trimbottom}{.02\height}
\renewcommand{\subfigurewidth}{0.192\linewidth}  

\begin{figure*}[h]
   \begin{subfigure}[l]{0.49\linewidth}
      \begin{subfigure}{\subfigurewidth}
         \adjincludegraphics[width=\textwidth, trim={{\trimleft} {\trimbottom} {\trimright} {\trimtop}},clip]{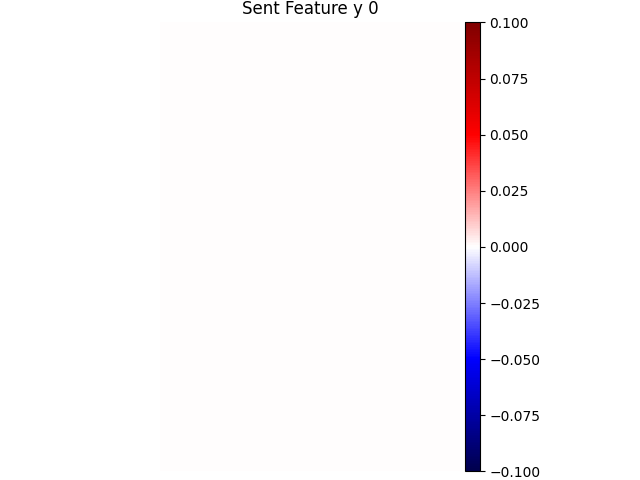}
         \caption*{$H \times W$}
      \end{subfigure}
      \begin{subfigure}{\subfigurewidth}
         \adjincludegraphics[width=\textwidth, trim={{\trimleft} {\trimbottom} {\trimright} {\trimtop}},clip]{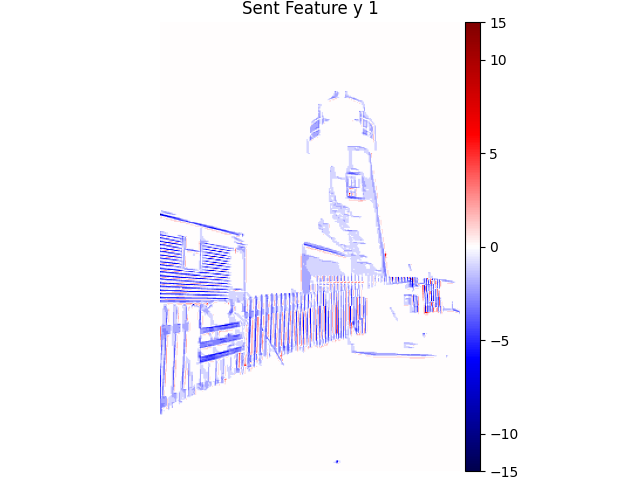}
         \caption*{$\frac{H}{2} \times \frac{W}{2}$}
      \end{subfigure}
      \begin{subfigure}{\subfigurewidth}
         \adjincludegraphics[width=\textwidth, trim={{\trimleft} {\trimbottom} {\trimright} {\trimtop}},clip]{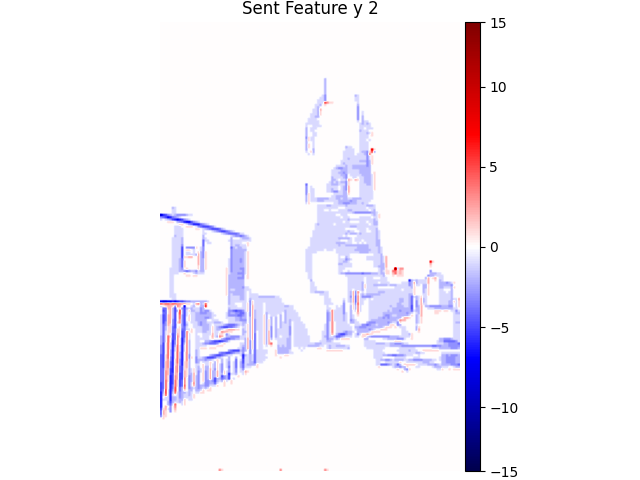}
         \caption*{$\frac{H}{4} \times \frac{W}{4}$}
      \end{subfigure}
      \begin{subfigure}{\subfigurewidth}
         \adjincludegraphics[width=\textwidth, trim={{\trimleft} {\trimbottom} {\trimright} {\trimtop}},clip]{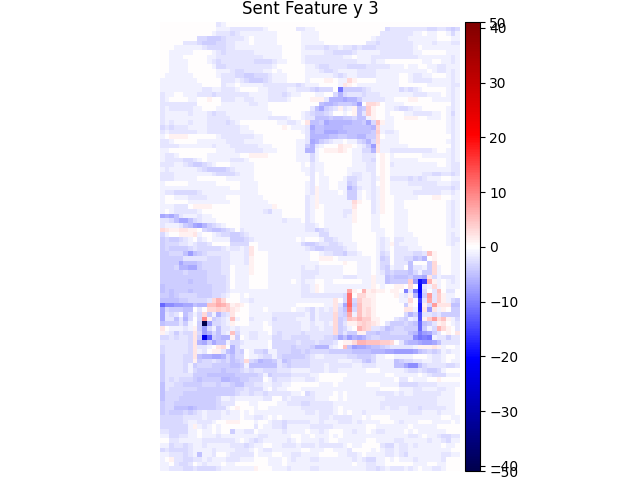}
         \caption*{$\frac{H}{8} \times \frac{W}{8}$}
      \end{subfigure}
      \begin{subfigure}{\subfigurewidth}
         \adjincludegraphics[width=\textwidth, trim={{\trimleft} {\trimbottom} {\trimright} {\trimtop}},clip]{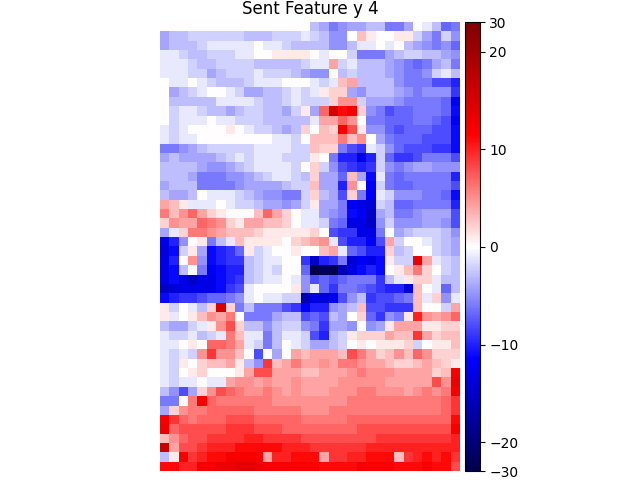}
         \caption*{$\frac{H}{16} \times \frac{W}{16}$}
      \end{subfigure}
      \caption{Rate is 0.145 bpp.}
   \end{subfigure}
   \hfill
   \begin{subfigure}[r]{0.49\linewidth}
      \begin{subfigure}{\subfigurewidth}
         \adjincludegraphics[width=\textwidth, trim={{\trimleft} {\trimbottom} {\trimright} {\trimtop}},clip]{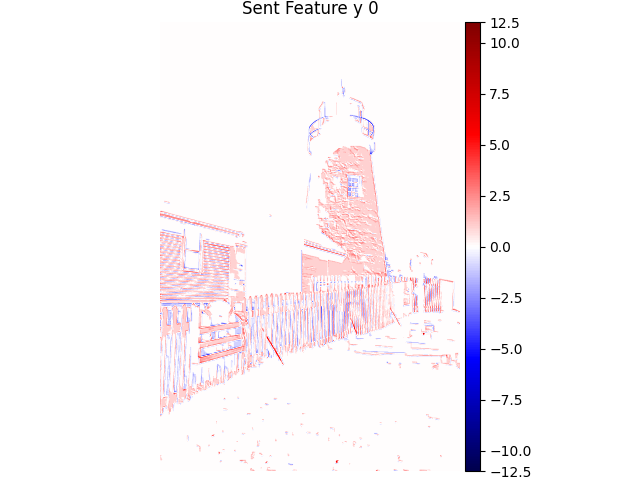}
         \caption*{$H \times W$}
      \end{subfigure}
      \begin{subfigure}{\subfigurewidth}
         \adjincludegraphics[width=\textwidth, trim={{\trimleft} {\trimbottom} {\trimright} {\trimtop}},clip]{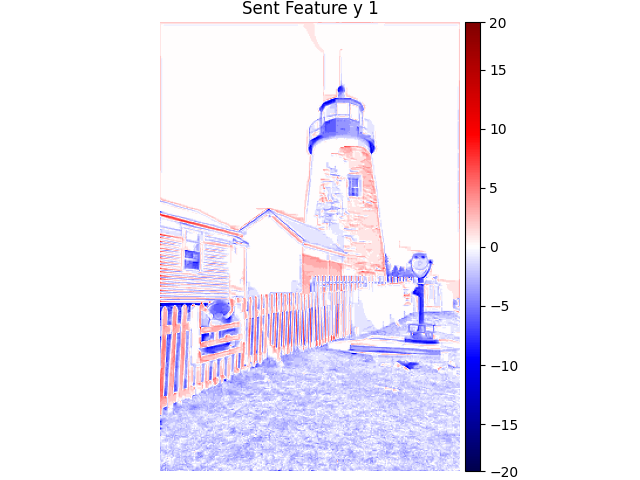}
         \caption*{$\frac{H}{2} \times \frac{W}{2}$}
      \end{subfigure}
      \begin{subfigure}{\subfigurewidth}
         \adjincludegraphics[width=\textwidth, trim={{\trimleft} {\trimbottom} {\trimright} {\trimtop}},clip]{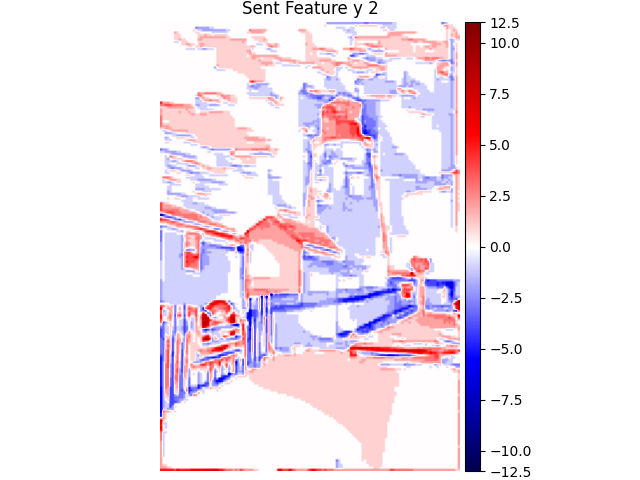}
         \caption*{$\frac{H}{4} \times \frac{W}{4}$}
      \end{subfigure}
      \begin{subfigure}{\subfigurewidth}
         \adjincludegraphics[width=\textwidth, trim={{\trimleft} {\trimbottom} {\trimright} {\trimtop}},clip]{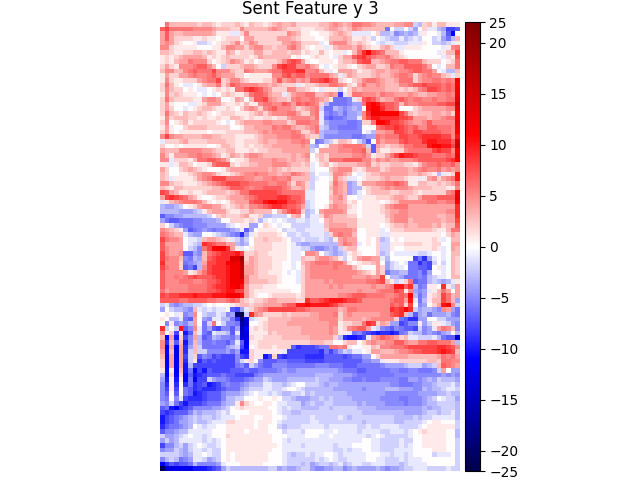}
         \caption*{$\frac{H}{8} \times \frac{W}{8}$}
      \end{subfigure}
      \begin{subfigure}{\subfigurewidth}
         \adjincludegraphics[width=\textwidth, trim={{\trimleft} {\trimbottom} {\trimright} {\trimtop}},clip]{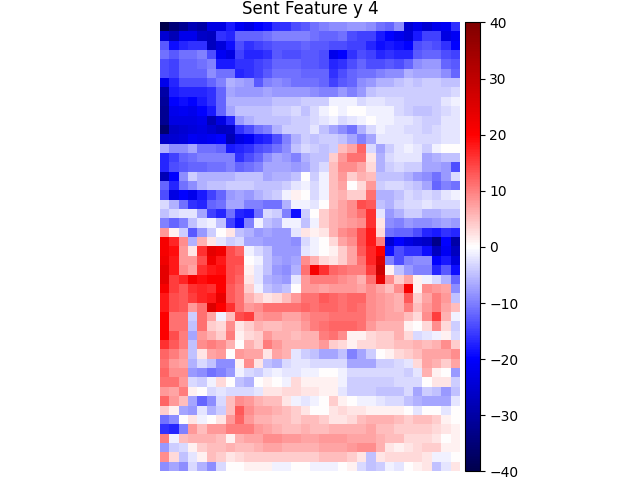}
         \caption*{$\frac{H}{16} \times \frac{W}{16}$}
      \end{subfigure}
      \caption{Rate is 0.589 bpp.}
   \end{subfigure}
   \caption{Latent variables for two rates (best viewed on screen). Captions indicate the resolution of each latent channel.}
   \label{fig:latent}
\end{figure*}

The main benefit of COOL-CHIC is the low complexity of its decoder. This is
illustrated in Fig. \ref{fig:complexity_learned_decoders}, which presents the
number of kMAC (kilo multiplication-accumulation) per decoded pixel for
different approaches. COOL-CHIC complexity remains consistently lower compared
with prior CNR-based approaches (e.g. COIN). Furthermore, it is two orders of
magnitude less complex than Ballé's hyperprior-based autoencoder while offering
competitive rate-distortion performance. This proves that COOL-CHIC is a
low-complexity alternative to learned autoencoders.
\newline

While the main focus of this work is the design of a low-complexity learned
decoder, it is possible to slightly increase the decoder complexity to obtain
better compression performance. In particular, the number of context pixels and
the width of the hidden layer are increased from 12 up to 24. Fig.
\ref{fig:bdrate_mac} presents the exploration of this performance-complexity
continuum. Performance is expressed as the BD-rate \cite{bdrate} of COOL-CHIC
using Ballé's hyperprior autoencoder as reference system. BD-rate indicates the
relative rate required to achieve the same quality than the reference system.
The point with the smallest complexity (680 MAC~/~decoded pixel) corresponds to
the one shown in the rate-distortion graphs of Fig. \ref{fig:rd_results} and
presents a BD-rate of 7~\% (i.e. COOL-CHIC needs 7~\% more rate to offer the
same quality than Ballé's autoencoder). Increasing the complexity up to 2~500
MAC~/~decoded pixel leads to decrease the BD-rate to less than 3~\%.

\subsubsection{Encoder complexity}

COOL-CHIC offers low decoding complexity. However, the encoding complexity is
more significant since COOL-CHIC learns the latent representation, the synthesis
MLP and the probability model MLP for each image. Here, the encoding of a $768
\times 512$ image takes around 10 minutes and 40~000 iterations. Yet, better
implementation (e.g. CUDA/C++) would lead to a dramatic speed-up. For instance,
Instant-NGP \cite{mueller2022instant} learns to synthesize images (albeit
without a rate constraint) in a matter of seconds. Besides reducing the duration
of each iteration, meta-learning-based approaches such as COIN++
\cite{DBLP:journals/corr/abs-2201-12904} offer solutions to significantly reduce
the number of iterations. This hints that the encoding time of COOL-CHIC is
likely not an issue.

The impact of the encoding time is evaluated by computing COOL-CHIC BD-rate
against different anchors (Ballé's autoencoder, COIN, HEVC and JPEG) throughout
the encoding process. The results of this experiment are presented in Fig.
\ref{fig:encoding_time_bd_rate}. COOL-CHIC requires less than 100 iterations
(1.5 second) to outperform JPEG and COIN, offering a BD-rate of -20~\% (versus
JPEG) and -30~\% (versus COIN). After 1 minute, COOL-CHIC already offers
compression performance close to Ballé's autoencoder with a BD-rate of 20~\%.
This results illustrates that the encoding time is not prohibitive since most of
the compression efficiency is obtained in a few minutes.
\newline

Conceptually, COOL-CHIC is better compared to conventional codecs (HEVC, VVC)
than to learned auto-encoders. Indeed, the encoding of COOL-CHIC consists of
learning  an adapted latent and transform which best suit the current image to
compress. This is similar to the trial of many different coding modes in
conventional codecs in order to find the best ones. It results in a high
encoding complexity (often done once on a dedicated server) and a low decoding
complexity (done many times on low-power devices e.g. smartphones).

\subsection{Visualization}

Visual examples are provided in Fig. \ref{fig:visual_comparison}, which compare
HEVC and COOL-CHIC reconstruction at different rates. At low rate, both codecs
exhibit significant degradations and compression artifacts. Since they operate
differently, the nature of their artifacts is different. HEVC is a block-based
codec and as such has blocking artifacts visible on the lighthouse wall or in
the grass. COOL-CHIC presents other kind of artifacts akin to banding artifacts
caused by the upscaling of low-resolution latent channels. At higher rates, both
HEVC and COOL-CHIC are able to produce high-quality reconstructions without coding
artifacts.
\newline

Unlike prior CNR-based codecs, COOL-CHIC does not rely on varying the MLPs
architecture to address different rates. Instead, the rate-distortion tradeoff
concerns mostly the latent representation, whose entropy decreases to reduce its
rate. This is depicted in Fig. \ref{fig:latent} which presents the 5
highest-resolution latent variables obtained for two different rates. The top
row in Fig. \ref{fig:latent} allows reconstruction of the image presented in
Fig. \ref{subfig:visualcomparison_low_rate} and the bottom row corresponds to
the image in Fig. \ref{subfig:visualcomparison_high_rate}. At high-rate
COOL-CHIC uses all latent resolutions, including the highest one, which is not
the case at low rate. Furthermore, low-rate latent channels are sparser and have
smaller entropy than the high-rate ones (e.g. the $\frac{H}{2} \times
\frac{W}{2}$ channel). This shows that the rate-distortion optimization of
COOL-CHIC learns to populate the latent variable according to the rate
constraint.

\section{Conclusion and future work}

This work proposes COOL-CHIC, a Coordinate-based Low Complexity Hierarchical
Image Codec. It is built on top of a Coordinate-based Neural Representation
(CNR), complemented by a hierarchical latent representation. By overfitting the
whole system for each image to compress, COOL-CHIC is able to achieve compelling
rate-distortion results for a reduced decoder-side complexity of 680
multiplications per decoded pixels. In particular, COOL-CHIC offers performance
on par with Ballé's hyperprior-based autoencoder while being two orders of
magnitude less complex. Moreover, COOL-CHIC performance comes close to modern
conventional codecs such as HEVC. This yields promising perspectives, as
the low decoder complexity paves the way for real-life usage of learned compression.
\newline

While image coding performance is already compelling, further progress is still
needed to compete with state-of-the-art codecs (VVC), especially at lower rates.
Also, studying how to extend COOL-CHIC to video coding is of primary interest,
as video compression requires low-complexity decoding in order to ensure
real-time decoding. Finally, reducing the encoding time is necessary to rely on
COOL-CHIC for practical use cases.

{\small
\bibliographystyle{ieee_fullname}
\bibliography{refs}

\begin{thebibliography}{10}\itemsep=-1pt

\bibitem{hevc}
Gary~J. Sullivan, Jens-Rainer Ohm, Woo-Jin Han, and Thomas Wiegand.
\newblock Overview of the high efficiency video coding ({HEVC}) standard.
\newblock {\em IEEE Transactions on Circuits and Systems for Video Technology},
  2012.

\bibitem{vvc}
Benjamin Bross, Jianle Chen, Jens-Rainer Ohm, Gary~J. Sullivan, and Ye-Kui
  Wang.
\newblock Developments in international video coding standardization after
  {AVC}, with an overview of versatile video coding ({VVC}).
\newblock {\em Proceedings of the IEEE}, 2021.

\bibitem{DBLP:conf/iclr/BalleMSHJ18}
Johannes Ball{\'{e}}, David Minnen, Saurabh Singh, Sung~Jin Hwang, and Nick
  Johnston.
\newblock Variational image compression with a scale hyperprior.
\newblock In {\em 6th International Conference on Learning Representations,
  {ICLR} 2018, Vancouver, BC, Canada, April 30 - May 3, 2018, Conference Track
  Proceedings}. OpenReview.net, 2018.

\bibitem{jpeg-ai-cfp}
{ISO/IEC JTC 1/SC29/WG1 N100250, REQ} "report on the {JPEG AI} call for
  proposals results", 2022.

\bibitem{Ma2022a}
Y. Ma, Y. Zhai, W. Jiang, I. Li, Z. Yang, and R. Wang.
\newblock {ROI} image codec optimized for visual quality.
\newblock 2022.

\bibitem{DBLP:conf/cvpr/HeYPMQW22}
Dailan He, Ziming Yang, Weikun Peng, Rui Ma, Hongwei Qin, and Yan Wang.
\newblock {ELIC:} efficient learned image compression with unevenly grouped
  space-channel contextual adaptive coding.
\newblock In {\em {IEEE/CVF} Conference on Computer Vision and Pattern
  Recognition, {CVPR} 2022, New Orleans, LA, USA, June 18-24, 2022}, pages
  5708--5717. {IEEE}, 2022.

\bibitem{DBLP:journals/corr/abs-2103-03123}
Emilien Dupont, Adam Golinski, Milad Alizadeh, Yee~Whye Teh, and Arnaud Doucet.
\newblock {COIN:} compression with implicit neural representations.
\newblock {\em CoRR}, abs/2103.03123, 2021.

\bibitem{mildenhall2020nerf}
Ben Mildenhall, Pratul~P. Srinivasan, Matthew Tancik, Jonathan~T. Barron, Ravi
  Ramamoorthi, and Ren Ng.
\newblock Nerf: Representing scenes as neural radiance fields for view
  synthesis.
\newblock In {\em ECCV}, 2020.

\bibitem{mueller2022instant}
Thomas M\"uller, Alex Evans, Christoph Schied, and Alexander Keller.
\newblock Instant neural graphics primitives with a multiresolution hash
  encoding.
\newblock {\em ACM Trans. Graph.}, 41(4):102:1--102:15, July 2022.

\bibitem{DBLP:journals/corr/abs-2210-06823}
Subin Kim, Sihyun Yu, Jaeho Lee, and Jinwoo Shin.
\newblock Scalable neural video representations with learnable positional
  features.
\newblock {\em CoRR}, abs/2210.06823, 2022.

\bibitem{DBLP:conf/iclr/BalleLS17}
Johannes Ball{\'{e}}, Valero Laparra, and Eero~P. Simoncelli.
\newblock End-to-end optimized image compression.
\newblock In {\em 5th International Conference on Learning Representations,
  {ICLR} 2017, Toulon, France, April 24-26, 2017, Conference Track
  Proceedings}. OpenReview.net, 2017.

\bibitem{DBLP:journals/corr/abs-1809-02736}
David Minnen, Johannes Ball{\'{e}}, and George Toderici.
\newblock Joint autoregressive and hierarchical priors for learned image
  compression.
\newblock {\em CoRR}, abs/1809.02736, 2018.

\bibitem{6327343}
Chi~Ching Chi, Mauricio Alvarez-Mesa, Ben Juurlink, Gordon Clare, Félix Henry,
  Stéphane Pateux, and Thomas Schierl.
\newblock Parallel scalability and efficiency of {HEVC} parallelization
  approaches.
\newblock {\em IEEE Transactions on Circuits and Systems for Video Technology},
  22(12):1827--1838, 2012.

\bibitem{begaint2020compressai}
Jean B{\'e}gaint, Fabien Racap{\'e}, Simon Feltman, and Akshay Pushparaja.
\newblock {CompressAI}: a {PyTorch} library and evaluation platform for
  end-to-end compression research.
\newblock {\em arXiv preprint arXiv:2011.03029}, 2020.

\bibitem{kodak}
{K}odak image dataset.
\newblock {\em http://r0k.us/graphics/kodak/}.

\bibitem{clic20pro}
Challenge on learned image coding 2020.
\newblock {\em http://clic.compression.cc/2021/tasks/index.html}.

\bibitem{DBLP:journals/corr/abs-2201-12904}
Emilien Dupont, Hrushikesh Loya, Milad Alizadeh, Adam Golinski, Yee~Whye Teh,
  and Arnaud Doucet.
\newblock {COIN++:} data agnostic neural compression.
\newblock {\em CoRR}, abs/2201.12904, 2022.

\bibitem{bdrate}
Gisle Bjontegaard.
\newblock Calculation of average psnr differences between rd-curves.
\newblock {\em VCEG-M33}, 2021.

\end{thebibliography}
}

\end{document}